\documentclass[12pt, a4paper]{article}
\pdfoutput=1
\usepackage{amsmath}
\usepackage{amsfonts}
\usepackage{amssymb}
\usepackage{graphicx, rotating}
\usepackage{epstopdf}
\usepackage{epsfig}
\usepackage{latexsym}
\usepackage{graphicx}
\usepackage{color}
\usepackage{amsmath,amssymb}
\usepackage{cite}
\usepackage{slashed}
\usepackage{hyperref}
\hypersetup{colorlinks, citecolor=bluscuro, linkcolor=black, urlcolor=bluscuro}
\definecolor{rossos}{cmyk}{0,1,1,0.55}
\definecolor{bluscuro}{rgb}{0.15, 0.2, .85}
\definecolor{bluchiaro}{cmyk}{1,.3,0.,0.1}

\newcommand{\eq}[1]{Eq.~(\ref{#1})}

\newcommand{\be}{\begin{equation}}
\newcommand{\ee}{\end{equation}}
\newcommand{\bea}{\begin{eqnarray}}
\newcommand{\eea}{\end{eqnarray}}
\newcommand{\bc}{\begin{center}}
\newcommand{\ec}{\end{center}}
\newcommand{\ba}{\begin{array}}
\newcommand{\ea}{\end{array}}

\def\lra#1{\overset{\text{\scriptsize$\leftrightarrow$}}{#1}}

\begin{document}

%FRONTPAGE2%%%%%%
\begin{titlepage}
\begin{flushright}
%UAB-FT--\\
August 2013
\end{flushright}
\vspace{.3in}

\begin{center}
\vspace{1cm}

{\Large \bf
%50 Closed Doors and 9 Open Windows in Physics Beyond the Standard Model
Towards the Ultimate SM Fit\\ 
\vskip.4cm
to Close in on Higgs Physics
}

\vspace{1.2cm}

{\large Alex Pomarol\,$^a$ and Francesco Riva\,$^{b}$\\}
\vspace{.5cm}
 {\it {$^a$\,Dept. de F\'isica, Universitat Aut{\`o}noma de Barcelona, 08193~Bellaterra,~Barcelona}}\\
{\it {$^b$\,Institut de Th\'eorie des Ph\'enom\`enes Physiques, EPFL,1015 Lausanne, Switzerland}}\\

\vspace{1.cm}

\begin{abstract}
\medskip
With the discovery of the Higgs at the LHC, experiments have finally addressed all aspects of the Standard Model (SM). At this stage, it is important to understand which windows for beyond the SM (BSM) physics are still open, and which are instead tightly closed. We address this question by parametrizing BSM effects with dimension-six operators and performing a global fit to the SM. We separate operators into different groups constrained at different levels, and provide independent bounds on their Wilson coefficients taking into account only the relevant experiments. Our analysis allows to assert in a model-independent way where BSM effects can appear in Higgs physics. In particular, we show that  deviations from the SM in the differential distributions
 of $h\to V\bar f f$  are related to other observables, such as triple gauge-boson couplings, and are then already constrained by present data. On the contrary, $BR(h\to Z\gamma)$ can still hide large deviations from the~SM.
\end{abstract}

\vspace{.4cm}

\end{center}
\vspace{.8cm}

\end{titlepage}

%%%%%%%%%%%%%%%%%%%%%%%%%%%%%%%%%%%%%%%%%%%%%%%%%%%%%%%%%%%%%%%%%%%%%%%%%%%%%%%%%%%%%%%%%%%%%%%%%%%%%%%%%%%%%%%%%%%%%%%%%%%%

\section{Motivation}\label{sec:0}

With the first LHC revenues we have begun  to gain experimental evidence for all sectors of the Standard Model (SM).
Not having  found    signs of  beyond the SM (BSM) physics, 
it is  time to ask  which  directions in the parameter space of   "deformations" of the SM  are  still unprobed.
This  can provide a useful guidance for future experiments.

The purpose of this paper   is to take steps towards   addressing  this  question.
We   will mainly focus   on   flavor-independent  observables,
and how these  constrain 
 electroweak symmetry breaking (EWSB)
effects in gauge-bosons and  Higgs physics.
As a  model-independent parametrization of  BSM physics, we
will use the Wilson coefficients of the independent dimension-six operators.
This is  a valid parametrization as  long as the BSM scale is heavier than the electroweak scale, as we will assume here.
Instead of an exhaustive fit to the SM, including all available data, 
our  goal is to show  which are the most relevant experiments that
 constrain the different directions in the parameter space of  Wilson coefficients,
and  provide the corresponding  bounds.

We will classify  the bounds  in different groups \cite{eemp2}.
First, those arising from $Z$-pole observables, $W$ mass and
some low-energy experiments
that  constrain,  at the per-mille level,  deviations in $W/Z$ propagators and gauge-boson couplings to fermions.
In a second  group, we  have bounds,  a factor $\sim 10$ weaker,
 from  measurements of the triple gauge-boson couplings (TGC),
that  will be crucial to  constrain certain directions in the parameter space, left unbound by the first group of experiments.
Finally, we will have bounds from  Higgs physics  arising from  the recent LHC data.

One of the  main purposes of our fit  is to
extract indirect constraints on  
possible deviations from the SM  in   future Higgs measurements.
In particular,  we will find that  only $\Gamma(h\to Z\gamma)$  can still hide large deviations from the SM prediction,  while deviations  in   $h\to V\bar f f$  are already constrained by other experiments,  as they can be related to TGC measurements. 
Since the LHC can  in principle do better measurements of TGC from processes such as $pp\to W\gamma,Z\gamma,WW,ZZ$, this will be a more optimal  way to constrain BSM physics than from 
$h\to V\bar f f$ or $VV\to h$.

There have been 
several   groups    also  addressing    these questions 
\cite{Masso:2012eq,Grinstein:2013vsa,Dumont:2013wma,Ciuchini:2013pca,Han:2004az,Contino:2013kra,delAguila:2011zs}. Contrary to ours,  however, 
most analyses have considered operators one by one, 
setting the rest  artificially  to zero,  and the combined effect of more than one operator 
 is rarely taken into account.
The analysis that goes closer in spirit to the one presented here is the one of Ref.~\cite{Han:2004az}.
  Our analysis differs from theirs in several aspects. 
   First of all, our analysis is extended to include Higgs physics and tries to understand where BSM effects
   can be found in future Higgs measurements. 
   Secondly, and  importantly,  our choice of basis~\cite{SILH,eemp1,eemp2} allows to separate  
   operators that are constrained at different level.
    This  choice minimizes the correlations between  Wilson-coefficient constraints
     and makes the results easy to interpret.
The analysis of Ref.~\cite{Han:2004az} was made 
 using  the basis of Ref.~\cite{Grzadkowski:2010es},
and 
led to  Ref.~\cite{Grinstein:2013vsa} to  erroneusly conclude that constraints from electroweak observables
were much weaker than what they actually are.

In Section~\ref{sec:1} we study the constraints on the  CP-even Wilson coefficients
arising from electroweak  (non-Higgs) observables.  
In Section~\ref{sec:Higgs} we  present  the relevant Wilson coefficients for Higgs physics,  show
their constraints, and study  new physics in  $h\to V\bar ff$.
In Section~\ref{sec:3} we briefly comment on CP-odd operators.
We leave for Appendix~\ref{app:LEP} the details concerning $Z$-pole observables, 
and in Appendix~\ref{app:cql3} we show how  LHC data can constrain  the relevant four-fermi interactions entering in our analysis.

%%%%%%%%%%%%%%%%%%%%%%%%%%%%%%%%%%%%%%%%%%%%%%%%%%%%%%%%%%%%%%%%%%%%%%%%%%%%%%%%%%%%%%%%%%%%%%%%%%%%%%%%%%%%%%%%%%%%%%%%%%%%

\section{Constraining BSM physics from EW observables}\label{sec:1}
%%%%%%%%%%%%%%%%%%%%%%%%%%%%%%%%%%%%%%%%%%%%%%%%%%
We will assume that  the BSM sector is heavy, with a mass scale  $\Lambda$ much larger than the 
weak-scale. This allows us to parametrize in a model-independent way all BSM effects
by   dimension-six operators added to the SM \cite{Buchmuller:1985jz,Grzadkowski:2010es}. 
There are many options for the choice of the dimension-6 operator basis. Here we will follow 
the basis of \cite{eemp1,eemp2} that we find more suitable for our
analysis than that of Ref.~\cite{Grzadkowski:2010es},
 due to its  better connection with  experiments. 
Assuming lepton and baryon number conservation, the full set of  operators in this basis can be found in  Tables 1 and 2 of Ref.~\cite{eemp2}: These contain  
 5 redundancies that we use  to  eliminate the 5 operators
 ${\cal O}_{2W,2B,2G}$ and  ${\cal O}^l_{L},{\cal O}^{(3)\, l}_{L}$. 
 
In Table~\ref{operatorsdim6} we show the dimension-6 operators of our basis that 
enter in  the observables   that we are interested in.
We  assume minimal flavor violation (MFV) \cite{D'Ambrosio:2002ex}, that is to say that the BSM sector respects a
$U(3)^5$ flavor-symmetry 
up to corrections proportional to the SM Yukawa couplings.  This allows us to concentrate on flavor-independent constraints. In particular, MFV implies that
dipole operator for fermions, as well as
$(i \tilde H^\dagger {\lra { D_\mu}} H)( \bar u_R\gamma^\mu d_R)$, 
are proportional to small Yukawa couplings and 
can  then be neglected.
In  the operators of Table~\ref{operatorsdim6}   we assume a contraction of  family indices  $i,j$ inside
each parenthesis, {\it e.g.}, $(\bar Q_L\gamma^\mu Q_L)=
(\bar Q_L^i\gamma^\mu Q_L^i)$, and $y_u\bar Q_L u_R=y_u^{ij}\bar Q_L^i u_R^j$, as implied by the MFV assumption at the leading order in a Yukawa expansion.
The top quark, having a large Yukawa coupling, could depart from the MFV assumption.
For this reason we will also  consider  the impact of treating  top operators separately.

CP-odd operators are not  included in Table~\ref{operatorsdim6},    since they do not interfere 
with the  SM contributions to the observables  that we are considering. 
Their Wilson coefficients only enter quadratically
in these processes and can  then be  neglected  at the linear level that we are working. 
In Section~\ref{sec:3}, however, we will briefly discuss their implications for TGC and $h\to V\bar f  f$.
\renewcommand{\arraystretch}{1.4}
\begin{table}[t]
\begin{center}
\begin{tabular}{|l|}\hline
${\cal O}_H=\frac{1}{2}(\partial^\mu |H|^2)^2$\\
${\cal O}_T=\frac{1}{2}\left (H^\dagger {\lra{D}_\mu} H\right)^2$\\
${\cal O}_6=\lambda |H|^6$
  \\ [.5ex]   \hline  
${\cal O}_W=\frac{ig}{2}\left( H^\dagger  \sigma^a \lra {D^\mu} H \right )D^\nu  W_{\mu \nu}^a$\\
${\cal O}_B=\frac{ig'}{2}\left( H^\dagger  \lra {D^\mu} H \right )\partial^\nu  B_{\mu \nu}$\\[.9ex]
\hline
 \end{tabular}\hspace{10mm}
 \begin{tabular}{|l|}\hline
${\cal O}_{BB}={g}^{\prime 2} |H|^2 B_{\mu\nu}B^{\mu\nu}$\\
${\cal O}_{GG}=g_s^2 |H|^2 G_{\mu\nu}^A G^{A\mu\nu}$\\
${\cal O}_{HW}=i g(D^\mu H)^\dagger\sigma^a(D^\nu H)W^a_{\mu\nu}$\\
${\cal O}_{HB}=i g'(D^\mu H)^\dagger(D^\nu H)B_{\mu\nu}$\\
${\cal O}_{3W}= \frac{1}{3!} g\epsilon_{abc}W^{a\, \nu}_{\mu}W^{b}_{\nu\rho}W^{c\, \rho\mu}$%\\
%${\cal O}_{3G}= \frac{1}{3!} g_s f_{ABC}G^{A\, \nu}_{\mu}G^{B}_{\nu\rho}G^{C\, \rho\mu}$
 \\[.5ex]\hline
 \end{tabular}
\end{center}
\begin{center}
\scalebox{0.94}{\tabcolsep=0.1cm\begin{tabular}{|l|l|l|}
\hline 
${\cal O}_{y_u}   =y_u |H|^2    \bar Q_L \widetilde{H} u_R$ + h.c.     &
${\cal O}_{y_d}   =y_d |H|^2    \bar Q_L Hd_R$ + h.c.       &
${\cal O}_{y_e}   =y_e |H|^2    \bar L_L H e_R$ + h.c.  
\\[.5ex]\hline
${\cal O}_{R}^u =
(i H^\dagger {\lra { D_\mu}} H)( \bar u_R\gamma^\mu u_R)$     &
${\cal O}_{R}^d =
(i H^\dagger {\lra { D_\mu}} H)( \bar d_R\gamma^\mu d_R)$       &
${\cal O}_R^e =
(i H^\dagger {\lra { D_\mu}} H)( \bar e_R\gamma^\mu e_R)$      
\\
${\cal O}_{L}^q=(i H^\dagger {\lra { D_\mu}} H)( \bar Q_L\gamma^\mu Q_L)$            & &
\\
${\cal O}_{L}^{(3)\, q}=(i H^\dagger \sigma^a {\lra { D_\mu}} H)( \bar Q_L\sigma^a\gamma^\mu Q_L)$     & 
$$            &    
 \\[.5ex]
 \hline
$\mathcal{O}_{LL}^{(3)\,ql}=( \bar Q_L \sigma^a\gamma_\mu Q_L)\, ( \bar L_L \sigma^a\gamma^\mu L_L)$
 &&
 $ \mathcal{O}_{LL}^{(3)\, l}=( \bar L_L \sigma^a\gamma^\mu L_L)\, (\bar L_L \sigma^a\gamma_\mu L_L)$
\\[.5ex]
 \hline
   \end{tabular}}
  \caption{\it Set of CP-even dimension-6 operators that defines our  basis. We are including only the four-fermion operators that affect our observables.  We  omit  dipole operators for fermions and  ${\cal O}_R^{ud}=y_u^\dagger y_d (i \tilde H^\dagger {\lra { D_\mu}} H)( \bar u_R\gamma^\mu d_R)$ since they are suppressed by light fermion Yukawas under the MFV assumption.
  Also ${\cal O}_{3G}$ is not included since it does not enter in our observables.
  The complete set of operators can  be found in Ref.~\cite{eemp2}.}\label{operatorsdim6} 
\end{center}
\end{table}

\subsection{Experimental input values}
We take the SM as defined by the   3 parameters $g$, $g'$ and $v\simeq 246$ GeV, that we relate
with the  well-measured values of  the Fermi constant $G_F$ as measured in muon decays, the fine-structure constant $\alpha_{\rm em}$, and the $Z$-boson mass $m_Z$.  New physics, parametrized through the dimension-6 operators of 
 Table~\ref{operatorsdim6},
 affects these 3 input observables. The subset of relevant operators is
 \begin{equation}
 \label{OpsInput}
%\Delta
\Delta\mathcal{L}_{\textrm{input}}=\frac{c_T}{v^2}\mathcal{O}_T+  \frac{c_V^+}{m_W^2} \left( \mathcal{O}_W+ \mathcal{O}_B\right)+\frac{c_{LL}^{(3)\, l}}{v^2}\mathcal{O}_{LL}^{(3)\, l}\, ,
\end{equation}
where we have defined 
\be
c_V^\pm=\frac{1}{2}(c_W\pm c_B)\, ,
\ee
and $c_{W,B}$ are the Wilson coefficients of the operators $\mathcal{O}_{W,B}$. 
 The orthogonal combination  $\mathcal{O}_W- \mathcal{O}_B$ does not affect the input parameters.
Notice that, instead of a common suppression scale $\Lambda$,  we are suppressing the operators of \eq{OpsInput} by the weak scale, either  $m_W$ or $v$, meaning that we have absorbed  the dependence on $\Lambda$  into
the Wilson coefficients. 
With this normalization the Wilson coefficients 
must  be considered  smaller than $O(1)$ for our expansion to make sense.

The modifications of the 3  input observables
due to the dimension-6 operators are given by
\begin{equation}\label{InputParamsShifts}
\frac{\delta \alpha_{\rm em}}{\alpha_{\rm em}}=-2s^2_{\theta_W} \hat S\ ,\ \ \  
\quad \frac{\delta m_Z^2}{m_Z^2}=  -\hat T+2s^2_{\theta_W}  \hat S\ ,\ \ \
\frac{\delta G_F}{G_F}=-2 c_{LL}^{(3)\, l}\,  ,
\end{equation}
where we have defined $s_{\theta_W}\equiv\sin\theta_W$ (and similarly for  other trigonometric functions), being
 $\theta_W$ the weak mixing angle, and  
\be
\hat S = 2{c_V^+}\ , \ \ \ \hat T= c_T\, ,
\ee
 characterize  the dominant contributions to the $W/Z$ propagators~\cite{Peskin:1990zt,Barbieri:2004qk}.

  Our fits for electroweak observables are performed using a $\chi^2$ analysis; the experimental data including correlations, are taken from Refs.~\cite{ALEPH:2005ab,LEP2wwzOLD}, while the SM predictions, including the dominant loop corrections, are taken from Refs.~\cite{Arbuzov:2005ma,Baak:2012kk} (we have checked that our results including  only $\hat S$ and $\hat T$   agree with Ref.~\cite{Baak:2012kk} at the percent level).
Henceforth  when  referring   to SM predictions to observables we will always be considering 
 predictions given  as a function  of these 3 input experimental values, $G_F$, $\alpha_{\rm em}$ and $m_Z$.

\subsection{$Z$-pole observables for leptons  and the  $W$ mass}
Among the most accurate experimental tests of the SM, stand the $Z$-pole observables measured at LEP-I/SLC, and the Tevatron measurement of the $W$ mass.  In a first set of constraints, we  single out  the leptonic observables
that are measured at the per-mille level. 
The relevant  dimension-6 operators that contribute to these quantities are,
apart from $\Delta\mathcal{L}_{\textrm{input}}$, 
\begin{gather}
\label{ObsLep1}
\Delta \mathcal{L}_{\textrm{leptons}}=\frac{c^e_{R}}{v^2}\mathcal{O}^e_{R}\, .
\end{gather}
It is important to notice that, being at the $Z$-pole,  four-fermion operators can be neglected.
LEP-I and SLC
measurements afford only  3 observables for the lepton sector,  
 that we can think of as  
$\Gamma_Z^{l_L}\equiv\Gamma(Z\to \bar l_L l_L)$, $\Gamma_Z^{l_R}\equiv\Gamma(Z\to \bar l_R l_R)$ 
 and  $\Gamma_Z^{\nu}\equiv\Gamma(Z\to \bar \nu\nu)$.  These can be extracted from the correlated set of observables $A_l,R_l,\sigma_{\rm had}^0,\Gamma_Z$ defined in \eq{LEPobservables} of Appendix~\ref{app:LEP}, 
 where  $\Gamma_Z^\nu=\Gamma_{Z}-\Gamma^{\rm visible}_{Z}$  since we  assume that there are no extra light degrees of freedom.
 The modifications  of these  quantities with respect to the  SM predictions 
due to  the  dimension-6 operators
 are given by  (see  Appendix~\ref{app:LEP} for details)
\begin{gather}\label{GammaLEP1}
{ \frac{\delta \Gamma_Z^\nu}{\Gamma_Z^\nu}}=\hat T-\frac{\delta G_F}{G_F}\ ,\quad
{ \frac{\delta \Gamma_Z^{l_L}}{\Gamma_Z^{l_L}}}= \frac{1}{ c^2_{2\theta_W}}\left(\hat T
-\frac{\delta G_F}{G_F}-4s^2_{\theta_W}\hat S\right)\ ,\\ 
{ \frac{\delta \Gamma_Z^{l_R}}{\Gamma_Z^{l_R}}}= -\frac{1}{c^2_{2\theta_W}}\left(\hat T-\frac{\delta G_F}{G_F}-2\hat S\right)-\frac{c_R^e}{s^2_{\theta_W}} \, .
\end{gather}
This gives 3 observables to constrain 4 Wilson coefficients. As  an additional observable 
needed to constrain all 4 coefficients,
we take   the $W$ mass, whose  correction from the SM value is given by
\begin{equation}
{ {\delta m_W \over m_W}}=\frac{1}{2 c_{2\theta_W}}\left[c^2_{\theta_W}\hat T
-s^2_{\theta_W}\left(\frac{\delta G_F}{G_F}+2 \hat S\right) \right]\, .
\end{equation}
Using this subset of observables, we can fit $c_T$, $c^+_V$, $c^e_R$ and $c_{LL}^{(3)\, l}$.
 Marginalizing over all but one coefficient at a time we find the 95\% C.L. intervals:
 \begin{gather}\label{boundsleptons}
c_T\in[-5,1]\times10^{-3}\, ,\quad\quad c_V^+\in[-6,0]\times10^{-3}\,,\\
c_R^e\in [-5,0]\times10^{-3}\,,\quad\quad c_{LL}^{(3)\, l}\in [-12,2]\times10^{-3}\, .\nonumber
\end{gather}
Other  four-lepton operators,  apart from ${\cal O}_{LL}^{(3)\, l}$, can  be constrained by LEP-II measurements 
of the cross-sections
$e^+e^-\to l^+l^-$ \cite{LEP2wwz}. These operators and their experimental constraints 
are completely orthogonal to our analysis and can be studied independently.  We will not pursue them further.

 %%%%%%%%%%%%%%%%%%%%%%%%%%%%%%%%%%%%%%%%%%%%%%%%%%%%%%%%%%%%%%%%%%%%%%%%%%%%%%%%%%%%%%%%%%%%%%%
 \subsection{Quarks}
 \noindent
Experimental data for quark physics abounds. The most accurate measurements come, again, from LEP-I physics at the $Z$-pole.
Assuming flavor-universality,
the relevant dimension-6 operators  for this type of physics are,
apart from $\Delta\mathcal{L}_\textrm{input}$, 
\begin{gather}
\label{OpsQuarks1}
\Delta \mathcal{L}_\textrm{quarks}= \frac{c^q_{L}}{v^2} \mathcal{O}^q_{L}+\frac{c_{L}^{(3)\, q}}{v^2}\mathcal{O}_{L}^{(3)\, q}+
\frac{c^u_{R}}{v^2} \mathcal{O}^u_{R}+\frac{c^d_{R}}{v^2} \mathcal{O}^d_{R}\, .
\end{gather}
\eq{OpsQuarks1} contains  4 new parameters that need 4 new observables in order to be constrained.
We take $R_b$, $\Gamma_Z^{\rm had}$, $A_b$ and $A_c$ as defined in  
Appendix~\ref{app:LEP}. They are affected by  dimension-6 operators as 
\bea
\frac{\delta \Gamma_Z^{\rm had}}{\Gamma_Z^{\rm had}}&\simeq&0.7 \,c_L^q +  2.3\,c_L^{(3)\, q}+ 0.3\, c_R^u - 0.2\, c_R^d -1.5\,\delta s_{\theta_W}^2+\hat T -\frac{\delta G_F}{G_F}\ ,\label{eqgammahad}\\
\frac{\delta R_b}{R_b}&\simeq&1.6 \,c_L^q - 0.05 \,c_L^{(3)\, q}- 0.3 \,c_R^u - 0.1 \,c_R^d +0.2\,\delta s_{\theta_W}^2\ ,\label{eqRb}\\
\frac{\delta A_c}{A_c}&\simeq& -0.9\, c_L^q + 0.9\, c_L^{(3)\, q} - 2.3\, c_R^u-4.2\,\delta s_{\theta_W}^2\ ,\label{eqAc}\\
\frac{\delta A_b}{A_b}&\simeq&0.1 \,c_L^q + 0.1\, c_L^{(3)\, q}+ 0.8\,c_R^d-0.6\,\delta s_{\theta_W}^2\, ,\label{eqAb}
\eea
where $\delta s^2_{\theta_W}$ is the correction to the effective weak mixing angle:
\begin{equation}
\delta s^2_{\theta_W}= \frac{1}{c_{2\theta_W}}\left[s^2_{\theta_W}\hat S -\frac{s^2_{2\theta_W}}{4}\left(\hat T-\frac{\delta G_F}{G_F}\right)\right]\, .
\label{deltas2W1}
\end{equation}
Only the first two observables, $\Gamma_Z^{\rm had}$ and  $R_b$, are measured at the per-mille level \cite{ALEPH:2005ab} and can give strong constraints on $c_L^{(3)\, q}$ and  
$c_L^q$,   since, accidentally, they enter respectively with  large coefficients   
in \eq{eqgammahad} and \eq{eqRb}.

The above  hadronic observables can be  complemented with information from  low-energy measurements.
One of the most relevant, due to its high accuracy, 
is the determination of the unitarity of the  CKM matrix by KLOE  and $\beta$-decays \cite{Antonelli:2010yf}.
This result can be interpreted 
as a measurement of the Fermi constant  in quark-lepton weak interactions, $G_F^{(q)}$, normalized to $G_F$ as extracted from $\mu$-decays:
\begin{equation}
\left(\frac{G_F^{(q)}}{G_F}\right)^2=
0.9999(12)\quad \textrm{at 95 \% C.L.}\, .
\label{GFq}
\end{equation}
Dimension-six operators modify this SM prediction: 
\begin{equation}
\delta\left(\frac{G_F^{(q)}}{G_F}\right)^2=4c_{LL}^{(3)\, l}+2c_{L}^{(3)\, q}-2c_{LL}^{(3)\, ql}\, ,
\label{deltagfq}
\end{equation}
which also includes a contribution from  4-fermion operators:
\be\label{opsq3}
\Delta \mathcal{L}_\textrm{quark-lepton}=\frac{c_{LL}^{(3)\, ql}}{v^2}\mathcal{O}_{LL}^{(3)\, ql}\, .
\ee
Now, using recent LHC data \cite{CMSHNC} for the measurement of the  high-energy differential cross-sections of 
$\bar q q\to \bar l \nu$, we can independently constrain  $c_{LL}^{(3)\, ql}$, 
 as shown in Appendix~\ref{app:cql3}. 
This allows us to  use \eq{deltagfq} to put
an independent constraint on   $c_{L}^{(3)\, q}$.
Putting all the above information  together and marginalizing (since the observables Eqs.~(\ref{eqgammahad})-(\ref{eqAb}) are also sensitive to $\Delta\mathcal{L}_\textrm{input}$, we must include also the leptonic operators and observables and perform a global fit: the resulting constraints on the coefficients of $\Delta\mathcal{L}_\textrm{input}$ will be given in the conclusion), we obtain the 95\% C.L. intervals:
\begin{gather}
c_{L}^{q}\in[-1,4]\times10^{-3}\,, \quad\quad  c_{L}^{(3)\, q} \in[-7,4]\times10^{-3}\,, \nonumber\\
c_{R}^{u}\in[-8,0]\times10^{-3}\,,\quad\quad c_{R}^{d} \in[-53,1]\times10^{-3}\,,\label{boundsquarks} \\
c_{LL}^{(3)\, ql}\in[-2,3]\times10^{-3}\, .\nonumber
\end{gather}

It is interesting to consider the case in which  operators made of the top quark
have different coefficients from  those associated to operators made with the first two families.
This is motivated by the  largeness of the  top Yukawa coupling 
that suggests that there can be large deviations from  flavor-universality. 
We then  add  the following operator to our analysis~\footnote{The other operators involving top-quarks, ($\mathcal{O}^{q_3}_{L}-\mathcal{O}_{L}^{(3)\, q_3})$ and  $\mathcal{O}^t_{R}$, affect only top couplings to gauge bosons and their constraints
arise from top physics at the LHC. We will not discuss them here.}:
\begin{gather}
\label{OpsQuarksMFV}
\Delta\mathcal{L}_\textrm{top}= \frac{c_{L}^{+\, q_3}}{v^2} \left(\mathcal{O}^{q_3}_{L}+\mathcal{O}_{L}^{(3)\, q_3}\right)\ ,\ \ \ \ 
{\rm where} \ \ c_{L}^{+\, q_3}=\frac{1}{2}(c_{L}^{q_3}+c_{L}^{(3)\, q_3})
\, ,
\end{gather}
that, due to the presence of  $b_L$ inside the 3rd family doublet $Q_L$,
gives an extra contribution to the    observables of Eqs.~(\ref{eqgammahad})-(\ref{eqAb}):
\begin{equation}
\left.\frac{\delta \Gamma^{\rm had}_Z}{\Gamma^{\rm had}_Z}\right|_{q_3}\simeq 0.5 \,c_{L}^{+\, q_3}\ ,
\quad\left.\frac{\delta R_b}{R_b}\right|_{q_3}\simeq 1.8 \,c_{L}^{+\, q_3}\ , \quad\quad\left.\frac{\delta A_b}{A_b}\right|_{q_3}\simeq 0.1\,c_{L}^{+\, q_3}\, .
\end{equation}
We have now 5 coefficients in the quark sector, \eq{OpsQuarks1} and \eq{OpsQuarksMFV}, 
but also 5 observables, those of 
Eqs.~(\ref{eqgammahad})-(\ref{eqAb}) and \eq{GFq}. KLOE data, which in the flavor-universal case was only used to slightly improve the numerical constraints, is now crucial to avoid unconstrained directions.
 The analysis gives:~\footnote{Introducing this extra parameter slightly alters the limits of Eqs.~(\ref{boundsleptons}) and (\ref{boundsquarks}), as we will show later.}
 \begin{gather}\label{topoperatore}
c_{L}^{+\, q_3}\in[-7,13]\times10^{-3}\,.
\end{gather}

Other  four-quark operators,  apart from \eq{deltagfq}, can  be constrained by LHC di-jet measurements \cite{Domenech:2012ai,deBlas:2013qqa}.
As for the lepton case,  these bounds do not interfere with our analysis, neither are relevant for Higgs physics; for this reason
they can  be studied in a separate context.

%%%%%%%%%%%%%%%%%%%%%%%%%%%%%%%%%%%%%%%%%%%%%%%%%%%%%%%%%%%%%%%%%%%%%%%%%%%%%%%%%%%%%%%%%%%%%%%%%%%%%%%%%%%%%%%%%%%%%%%%%%%%
\subsection{Triple gauge-boson couplings}\label{sec:TGC}

Extra important experimental data needed to further constrain our set of operators is 
the measure of the $ZWW$ and $\gamma WW$ couplings. The best measurements still come from $e^+e^-\to W^+W^-$ at LEP-II,
although LHC data is starting to be competitive.
The dimension-6 operators involved in this process 
 are, apart from Eqs.~(\ref{OpsInput}) and (\ref{ObsLep1}),
\begin{equation}
\Delta\mathcal{L}_\textrm{TGC}=
\frac{\kappa_{HV}^+}{m_W^2} \left(\mathcal{O}_{HW} +\mathcal{O}_{HB}\right)+
\frac{c_V^-+\kappa_{HV}^-}{2m_W^2} \mathcal{O}_{+}
+\frac{\kappa_{3W}}{m_W^2} \mathcal{O}_{3W}\, ,
\label{tgc}
\end{equation}
where we have defined 
\be
\kappa_{HV}^\pm=\frac{1}{2}(\kappa_{HW}\pm\kappa_{HB})\, ,
\ee
with $\kappa_{HW,HB}$ the Wilson coefficients of $\mathcal{O}_{HW,HB}$, and
\be
\mathcal{O}_{\pm}\equiv  (\mathcal{O}_W- \mathcal{O}_B)\pm
(\mathcal{O}_{HW}-\mathcal{O}_{HB})\, .
\ee
Notice that in order to connect with the different experiments, we are, for convenience, working  with 
the orthogonal combinations 
$\{\mathcal{O}_W+ \mathcal{O}_B,\mathcal{O}_{HW}+\mathcal{O}_{HB},
\mathcal{O}_{\pm}\}$
instead of the original subset $\{\mathcal{O}_W,
\mathcal{O}_B,\mathcal{O}_{HW},\mathcal{O}_{HB}\}$ given in  Table~\ref{operatorsdim6}.
It is easy to show that the coefficients of $\mathcal{O}_{\pm}$ are given by the combinations $(c_V^-\pm\kappa_{HV}^-)/2$.
In \eq{tgc} we have not included the  operator  $\mathcal{O}_{-}$  
since it does not enter in  TGC,  and it is only relevant for Higgs physics  as we will see later.
Indeed, one finds 
  the following contributions to the  $ZWW$ and $\gamma WW$ vertices~\cite{Hagiwara:1986vm}:
\bea
\Delta {\cal L}_{3V} &=& i g\delta g_1^Z c_{\theta_W}
 Z^{\mu}  \left( W^{- \, \nu}  W^+_{\mu\nu} 
  - W^{+ \, \nu}   W^-_{\mu\nu} \right) 
 \nonumber 
+
ig \left(  \delta \kappa_Z c_{\theta_W} Z^{\mu\nu} + \delta\kappa_\gamma s_{\theta_W} A^{\mu\nu}\right) 
W^-_\mu  W^+_\nu
\nonumber \\
&&+
\frac{ig}{m^2_W}\left(  \lambda_Z c_{\theta_W} Z^{\mu\nu} + \lambda_\gamma s_{\theta_W} A^{\mu\nu}\right) W^{- \rho}_{\nu}  W^{+}_{\rho \mu}\ ,
\label{V3}
\eea
where   $ V_{\mu\nu} \equiv  \partial_\mu V_\nu -  \partial_\nu V_\mu$ for $V=W^\pm,Z,A$, and
\bea
 \delta g_1^Z &=&\frac{c_W+\kappa_{HW}}{c^2_{\theta_W}}=
  \frac{1}{c^2_{\theta_W}} 
 \left( c_{V}^+ + c_V^-+\kappa_{HV}^- + \kappa_{HV}^+\right ) \ ,  \nonumber\\
  \delta \kappa_\gamma &=&\kappa_{HW}+\kappa_{HB}= 
2 \kappa^+_{HV}\ ,\qquad\quad 
 \delta \kappa_Z= \delta g_1^Z - t^2_{\theta_W}  \delta \kappa_\gamma
\ \nonumber  ,\\
  \lambda_Z &=&  \lambda_\gamma =  \kappa_{3W} \, .
\label{3vert}
\eea
Notice that  out of the 5  quantities in \eq{V3},
only 3 are independent if one considers only dimension-6 operators;  we take these to be $\delta g_1^Z$, $\delta \kappa_\gamma$ and 
$\lambda_\gamma$.
This means that data from the differential cross-sections  of $e^+e^-\to W^+W^-$
 can only constrain 3 extra combinations of Wilson coefficients
 that corresponds to  those of \eq{tgc}:
$\kappa_{HV}^+$, $c_V^-+\kappa_{HV}^-$ and $\kappa_{3W}$.\footnote{Other Wilson coefficients entering
 in the process $e^+e^-\to W^+W^-$, such as those of Eqs.~(\ref{OpsInput}) and (\ref{ObsLep1}), 
 can be neglected since their  constraints, derived in our previous analysis, are 
stronger   than the ones we will be obtaining here.}
 Unfortunately a  three-parameter analysis has not been provided by the full LEP-II collaboration \cite{LEP2wwzOLD,LEP2wwz}, but only by DELPHI  \cite{Abdallah:2010zj}.
Nonetheless, a small value  $\kappa_{3W}$ is expected in most theories where the  SM gauge bosons are assumed to be elementary above  the BSM scale $\Lambda$. Neglecting  this contribution  (and then taking $\lambda_\gamma=0$), we can 
use the two-parameter fit of $g_1^Z$ and $\kappa_\gamma$ from Ref.~\cite{LEP2wwzOLD}, to obtain
the  95\% C.L. intervals:
\begin{equation}
c_V^-+\kappa_{HV}^- \in[-4.4,6.6]\times10^{-2}\,,\quad \kappa_{HV}^+  \in[-5.5,3.9]\times10^{-2}\,.
\label{tgcbounds}
\end{equation}
Including in the fit $\kappa_{3W}$  does not  change 
considerably our results; using DELPHI data \cite{Abdallah:2010zj} (which is however strongly correlated with Ref.~\cite{LEP2wwzOLD}) one finds results similar to \eq{tgcbounds} and $\kappa_{3W} \in[-4,7]\times10^{-2}$.

%%%%%%%%%%%%%%%%%%%%%%%%%%%%%%%%%%%%%%%%%%%%%%%%%%%%%%%%%%%%%%%%%%%%%%%%%%%%%%%%%%%%%%%%%%%%%%%%%%%%%%%%%%%%%%%%%%%%%%%%%%%%
\section{Higgs physics}\label{sec:Higgs}

We now extend our analysis to Higgs physics. 
Apart from the operators already introduced before that could also affect Higgs physics, we have  8   CP-even dimension-six operators that give contributions only to Higgs physics and not to other SM processes. These are those operators that can be built from the Higgs modulus, {\it i.e.}, $|H|^2$.
In our basis,  Table~\ref{operatorsdim6}, these operators are
\begin{equation}\label{higgsop}
\Delta\mathcal{L}_\textrm{Higgs}=
\frac{c_H}{v^2}\mathcal{O}_H+\sum_{f=t,b,\tau}\frac{c_{y_f} }{v^2}  \mathcal{O}_{y_f}+\frac{c_6 }{v^2} \mathcal{O}_6
+
\frac{}{} \frac{\kappa_{BB} }{m_W^2} \mathcal{O}_{BB}+\frac{\kappa_{GG}}{m_W^2}\mathcal{O}_{GG}
+\frac{c_V^--\kappa_{HV}^-}{2m_W^2} \mathcal{O}_{-}\, ,
\end{equation}
which holds for one family, taken to be the 3rd one as it is the most relevant for Higgs physics.
The  presence of  $\mathcal{O}_{-}$ must be traced  back \cite{eemp2} to the fact that the operator ${\cal O}_{WW}=g^2|H|^2W^{\mu\nu\, a} W_{\mu\nu}^a$, that can obviously   affect only Higgs physics,  when
 written in our basis is given by 
 \be
 {\cal O}_{WW}=4\mathcal{O}_{-}+{\cal O}_{BB}\, .
 \ee
Now, the operator ${\cal O}_6$ is in one-to-one correspondence with a deviation in the triple-Higgs coupling that has not yet been measured at the LHC. Therefore, we will not be considering it any longer.
The 7 remaining operators of \eq{higgsop}  affect  the main  Higgs  branching ratios, as well as Higgs production 
cross-sections at the LHC. In particular,
we have that the 4 Wilson coefficients $c_{y_b}$, $c_{y_\tau}$, $\kappa_{BB}$, $\kappa_{HV}^-$
enter respectively in $BR(h\to bb)$, $BR(h\to \tau\tau)$, $BR(h\to \gamma\gamma)$, $BR(h\to Z\gamma)$, while
$\kappa_{GG}$ affects $\sigma(GG\to h)$;
the coefficient $c_H$ gives a contribution to the Higgs propagator and then enters universally in all Higgs processes (affecting, in particular, vector-boson fusion),
while $c_{y_t}$ modifies the $h\bar tt$ coupling 
  that  enters indirectly in  $BR(h\to \gamma\gamma)$ and $BR(h\to Z\gamma)$ at the one-loop level \cite{SILH},
and also enters in the associated  production process $pp\to h\bar tt$ that the 
LHC can  become sensitive to in the near  future.
 The Lagrangian for these processes is determined by
\be
 {\cal L}_{h}^\textrm{SM}+\Delta{\cal L}_{h}\, ,
 \label{LH0}
\ee
where
\begin{equation}\label{LSM}
{\cal L}_{h}^\textrm{SM}=\frac{1}{2}(\partial_\mu h)^2+(\sqrt{2}G_F)^{1/2}\, h\Big[-m_f (\bar f_L f_R + {\rm h.c.})+m_Z^2
Z_\mu Z^\mu+2m_W^2W_\mu^+W^{-\,\mu}\Big]+\cdots\, ,
\end{equation}
gives the SM contribution to  single-Higgs processes written as a function of the physical masses, $m_f$, $m_Z$ and 
 $m_W^2=m_Z^2(1+\sqrt{1-2^{3/2}\pi\alpha_{\rm em} /m_Z^2G_F})/2$,
while
\begin{align}\label{deltaL1}
\Delta{\cal L}_{h} =\frac{c_H}{2} (\partial^\mu h)^2\, &+\frac{h}{v}\Big[ \delta g_{hff}(\bar f_L f_R + {\rm h.c.})
+\delta g_{hZZ}   Z_\mu Z^\mu 
+\delta g_{hWW} W_\mu^+ W^{-\,\mu} \nonumber\\
& +4\kappa_{BB}s_{\theta_W}^2 A_{\mu\nu}A^{\mu\nu}+4\kappa_{GG}\frac{g_s^2}{g^2}\, G_{\mu\nu}^AG^{\mu\nu\, A}+4t_{\theta_W}\kappa_{Z\gamma}Z_{\mu\nu}A^{\mu\nu}\Big]\, ,
\end{align}
gives the contributions proportional to the Wilson coefficients  of $\Delta{\cal L}_{\rm input}$, $\Delta{\cal L}_{\rm Higgs}$, and  also that of ${\cal O}_+$ in \eq{tgc} that enters in $hZ\gamma$.
The first line of \eq{deltaL1} includes the corrections to the SM  $h\bar f f$ and $hVV$   interactions ($V=W,Z$) due to the   Wilson coefficients entering  in the input parameters,  in the wave-function renormalization of  $V$
and  in direct contributions arising from $\mathcal{O}_T$.  
They read
\begin{equation}\label{shiftHZZ}
\frac{\delta g_{hZZ}}{g_{hZZ}^{\rm SM}}=- c_T    -\frac{\delta G_F}{\, 2G_F}\ , \quad\quad \frac{\delta g_{hWW}}{g_{hWW}^{\rm SM}}= 2 \, \frac{\delta m_W}{m_W} -\frac{\delta G_F}{\, 2G_F} \  ,\quad \quad  \frac{\delta g_{hff}}{g_{hff}^{\rm SM}}=-c_{y_f}-\frac{\delta G_F}{2G_F} \, .
\end{equation}
With the exception of  $c_{y_f}$, we know from the analysis of Section~\ref{sec:1} that these effects are  small.
The second line of \eq{deltaL1}  gives the corrections to  $h\gamma\gamma$, $hGG$ and 
$hZ\gamma$ coupling, where
\begin{equation}
\kappa_{Z\gamma}=- \frac{1}{2}\kappa^-_{HV}-2s^2_{\theta_W}\kappa_{BB}\, .
\label{hzg}
\end{equation}

The dominant modifications from \eq{deltaL1}
to  $BR(h\to \bar f f)$, $BR(h\to \gamma\gamma)$, $BR(h\to Z\gamma)$ and  $\sigma(GG\to h)$
  can be found in \cite{SILH}.
We confront these with a combination of ATLAS \cite{ATLAS:2013sla} and CMS \cite{CMS:ril,CMS:xwa,CMS:bxa} data (for technical details see Ref.~\cite{Couplings}). 
For this fit
we also include the measurements of $pp\to h\to VV^*$  and vector-boson fusion
$pp\to  VV qq\to h qq$; there are  extra contributions
to   these processes, as we will see in the next Subsection (Eqs.~(\ref{LhVV}) and (\ref{LhVff})),
but these, we can advance,
are found to be small and can be neglected here.
Strong bounds can be found for  Wilson coefficients entering at tree-level
 in the  effective
  Higgs couplings to $GG$, $\gamma\gamma$ and $Z\gamma$, 
as they  arise in the SM  at the one-loop level. 
 Indeed, marginalizing over $c_H$ and $c_{y_f}$,   we obtain at the 95\%~C.L.
 \begin{gather}
\kappa_{GG}\in[-0.8,0.8]\times10^{-3}\,,\quad \kappa_{BB}\in[-1.3,1.8]\times10^{-3}\,,\quad \kappa_{Z\gamma}\in[-6,12]\times10^{-3}\, .
\end{gather}
Notice that the constraints on $g_S^2 \,\kappa_{GG} $ are comparable to those on ${g'}^2\kappa_{BB} $, and that the bound on the $hZ\gamma$  coupling is quite strong even though the experimental data is only sensitive to 
$BR(h\to Z\gamma)$ that are $\sim 10$ above  the SM value \cite{Chatrchyan:2013vaa,ATLAS:2013rma}.
For the Wilson coefficients $c_H$ and $c_{y_f}$
we do not find significant bounds   after
marginalizing over the others (for this reason  we do not report them).
Although this implies that   there is   room for BSM here,
we recall that  these  coefficients cannot be larger than one if we want our expansion (in higher-dimensional operators)
to be reliable.

This ends  our  analysis for  getting the most important constraints on the   operators of Table~\ref{operatorsdim6}. 
Our results can be particularly useful to  bound  possible new physics effects in future   new Higgs
measurements.
We show this below with the example of the  $h\to V\bar f f$  decay.

%%%%%%%%%%%%%%%%%%%%%%%%%%%%%%%%%%%%%%%%%%%%%%%%%%%%%%%%%%%%%%%%%%%%%%%%%%%%%%%%%%%%%%%%%%%%%%%%%%%%%%%%%%%%%%%%%%%%%%%%%%%%%%%%%%%%%%%%%%%%%%%%%%%%%%
\subsection{New physics effects  in  $h\rightarrow V\bar ff$}\label{sec:hVff}

 The decays  $h\to V\bar f f$ ($V=W,Z$)  
  are potentially much richer than  two-body decays, since the different
 differential partial-widths can give in principle extra information on  BSM contributions~\cite{Choi:2002jk,Cao:2009ah,Gao:2010qx,Stolarski:2012ps,Isidori:2013cla,Isidori:2013cga,Grinstein:2013vsa}.
Nevertheless, as we will show,  most of   the new information that we could extract from measuring the various
differential partial-widths of the decay $h\rightarrow V\bar ff$  is already  constrained by other experiments.

Contributions  to 
$h\rightarrow V\bar ff$   can come  from   corrections to 
$hVV$ vertices and contact-interactions  $hV\bar ff$.
Apart from the contributions given in \eq{LH0}, we have
\bea
 \Delta{\cal L}_{hVV} &=& 2\frac{h}{v} \Big[\hat c_{W} \, \left(  W^-_\mu {\cal D}^{\mu \nu} W^+_\nu +{\rm h.c.} \right)
+ \hat c_{Z}\,Z_\mu {\cal D}^{\mu \nu} Z_\nu +(\hat c_{W}-\hat c_B)t_{\theta_W}\, Z_\mu {\cal D}^{\mu \nu} A_\nu \Big]\nonumber  \\
& -&2 \, \frac{h}{v} \, \Big[c_{WW} \, W^{+\mu \nu} W^-_{\mu \nu} +c_{ZZ} \, Z^{\mu \nu}Z_{\mu \nu}
\Big]\ ,\label{LhVV}\\
\Delta{\cal L}_{hVff} &=&\frac{h}{v}\sum_{f=f_L,f_R}\Big[
 g_{hWf f'}\, W_\mu \bar f \gamma^\mu f'+g_{hZf f} \,  Z_\mu \bar f\gamma^\mu f\Big]\ , \label{LhVff}
\eea
where ${\cal D}_{\mu \nu}= \partial_\mu\partial_\nu-\Box \eta_{\mu\nu}$ and 
\bea
&&\hat c_{W}= c_W+\kappa_{HW}\ , \qquad
\hat c_{Z}=\hat c_{W}+ \hat c_B t^2_{\theta_W} 
\ , \qquad \hat c_{B}=c_B+\kappa_{HB}\, ,\\
&&
c_{WW}=\, \kappa_{HW}\ ,\qquad
c_{ZZ}=\frac{1}{2} (\kappa_{HW} +\kappa_{HB} t^2_{\theta_W})-2 \frac{s_{\theta_W}^4}{c_{\theta_W}^2} \kappa_{BB}\, .\label{cwwczz}
\eea
\eq{LhVff} gives the contributions to the  contact $hV\bar f f$ vertices 
that is found to be correlated with  those   to  the $V\bar ff$  vertices:
\be
g_{hZff}=\frac{2}{v}\delta g_Z^f \quad \textrm{and}\quad g_{hWff'}=\frac{2}{v}\delta g_W^f\, ,
\label{coups}
\ee
where $\delta g_Z^f$ and $\delta g_W^f$ are given respectively in  Eqs.~(\ref{shifts}) and (\ref{deltashiftW})
of  Appendix~\ref{app:LEP}.

The CP-even part of the total amplitude for the process $h\to V\bar f f$ can be written as~\footnote{We neglect terms proportional to the light fermion masses (see however Ref.~\cite{Isidori:2013cla}). Also we omit a term
proportional to $C^W_f\epsilon_{\mu\nu\alpha\beta}\, p^{\alpha}\, q^{\beta}$ that could be CP-even if $C^W_f$ is pure imaginary. None of the Wilson coefficients of the dimension-6 operators contribute to this term at tree-level.} 
\begin{align}
 \mathcal{M}(h \to VJ_{f}) = \,  (\sqrt{2} G_F)^{1/2} \epsilon^{*\mu}(q) \,  J^{V\, \nu}_{f}(p) \left[ \, A^V_{f}  \, \eta_{\mu\nu} \, + \, B^V_{f} \,\left( p\cdot q\, \eta_{\mu\nu} -p_{\mu}\, q_{\nu}\right)  \right] \, ,
 \end{align}
where $q$  and $p$ are respectively  the total 4-momentum of  $V$ and 
the fermion pair in the $J^V_{f}$ current ($J_{f_{L,R}}^\mu=\bar f_{L,R}\gamma^\mu f_{L,R}$),
$\epsilon^\mu$ is the polarization 4-vector of $V$,
and we have defined
\begin{align}
A^V_{f}  = & \,a^V_{f}+\widehat a^V_{f}\frac{  p^2+m_V^2}{p^2-m_V^2}\, ,
& B^V_{f} = & \, b^V_{f}\frac{1}{p^2-m_V^2} +\widehat b^V_{f}\frac{1}{p^2}\ \ \ \ \ \ (\widehat b^V_f=0\ {\rm for}\ V=W ) \,.
\end{align}
The above coefficients are in one-to-one correspondence with the coefficients of the  Lagrangians Eqs.~(\ref{deltaL1}), (\ref{LhVV}) and (\ref{LhVff}):
\begin{align}
a^Z_{f}&=-g_Z^{f}(1+\frac{\delta g_{hZZ}}{m_Z^2}) +2 eQ_f   (\hat c_{W}-\hat c_B) t_{\theta_W}  +v g_{hZff} \, ,&
 a^{W}_{f} &=-g_W^f(1+\frac{\delta g_{hWW}}{2m_W^2})  +v g_{hWff^\prime}\, ,\nonumber \\
\widehat a^Z_{f}&=g_Z^{f}(1+\frac{\delta g_{hZZ}}{m_Z^2}+2 \hat c_{Z})\, , & 
\widehat a^{W}_{f}& =g_W^f(1+\frac{\delta g_{hWW}}{2m_W^2}+2 \hat c_W ) \, ,  \nonumber\\
b^Z_{f}&=8 g_Z^{{f}}  c_{ZZ}  \, , &
b^{W}_{f}& =4 g_W^{f}  c_{WW} \, ,\nonumber \\
\widehat b^Z_{f}&=-8 eQ_f t_{\theta_W}\kappa_{Z\gamma}\, ,   &         
\label{asbs}
\end{align}
where we have not included  the universal contribution from $c_H$ that  drops when calculating $BR$. 
All contributions to the $7$ quantities in \eq{asbs} correspond to Wilson coefficients that have already been constrained by other experiments.
Indeed, the terms proportional to $\delta g_{hVV}$ of \eq{shiftHZZ} and the universal part of the shifts in $g_V^{f}$, given in  Eqs.~(\ref{shiftUNI}) and (\ref{deltashiftW}), are constrained, as we have shown, at the per-mil level, 
and can be readily neglected.
Similarly,  $\kappa_{BB}$ that appears in  \eq{cwwczz} is constrained from $BR(h\to \gamma\gamma)$.
The contributions proportional to $g_{hZff}$ that depends, in particular, on $c^{u,d}_R$,
are also constrained but, as shown in \eq{boundsquarks}, only at the per-cent level.
Nevertheless, 
their effects
in total amplitudes squared, when summed over the different  flavors, can be constrained much more.
This is due to the interesting relation
\begin{equation}\label{formulaalex}
\sum_{f=quarks} g_Z^f\, g_{hZff}=\frac{2}{v}\sum_{f=quarks} g_Z^f \, \delta g_Z^f=\frac{1}{v} \frac{\delta \Gamma_Z^{\rm had}}{\Gamma_Z^{\rm had}}\sum_{f=quarks}  (g_Z^{f})^2+O(\hat S, \hat T,\delta G_F/G_F)\ ,
\end{equation}
where in the first equality we have used \eq{coups} and in the second  we have used the hadronic part of \eq{shiftGammaZtot} that is true up to pieces proportional to $\hat S, \hat T$ and $\delta G_F/G_F$. Since all quantities on the right-hand side of \eq{formulaalex}
are very well measured (per-mille level), we obtain  a constraint on the effect of 
$g_{hZff}$ at this level. Neglecting these terms, 
we are left with only  3  Wilson coefficients, $c_V^-$ and  $\kappa_{HV}^\pm$, that are constrained by the measurements of the $Z(\gamma)WW$ 
and $hZ\gamma$ couplings and whose   bounds are not so strong.
Using \eq{3vert} and \eq{hzg}, we can relate the $7$ coefficients of \eq{asbs} with the $3$ experimental values of $\delta g_1^Z$, $\delta \kappa_\gamma$ and 
 $\kappa_{Z\gamma}$. The result is
\begin{align}
a^Z_{f}&\simeq-g_Z^{f,\rm SM}+2eQ_f\left(s_{2\theta_W}\delta g^Z_1 - t_{\theta_W}\delta \kappa_\gamma \right),&       
a^{W}_{f} &\simeq -g_W^{f,\rm SM}\, ,\nonumber \\
\widehat a^Z_{f}&\simeq g_Z^{f,\rm SM}(1+2c_{2 \theta_W}\delta g^Z_1  + 2 t^2_{\theta_W}\delta \kappa_\gamma) \, , &       
\widehat a^{W}_{f}& =g_W^{f,\rm SM}(1+2c^2_{\theta_W}\delta g_1^Z) \, ,\nonumber\\
b^Z_{f}&\simeq  
\frac{2g_Z^{f,\rm SM}}{c^2_{\theta_W}} \left(\delta  \kappa_\gamma-4c_{2\theta_W} \kappa_{Z\gamma}  \right)   \, , &       
 b^{W}_{f}& \simeq 2g_W^{f,\rm SM} \left(\delta \kappa_\gamma-4 \kappa_{Z\gamma}\right) \, ,\nonumber \\
\widehat b^Z_{f}&\simeq-8e Q_ft_{\theta_W}\kappa_{Z\gamma}\, .   &        
  \label{asbsobs}
\end{align}
We can use the experimental constraints on  
$\delta  g_1^Z$, $\delta  \kappa_\gamma$ and 
 $\kappa_{Z\gamma}$ to  put a constraint on the size of these  7 quantities.         
 For example, for the  case of $h\to Z \bar ll$, we find 
\begin{align}
&\frac{\delta a^Z_{l_L}}{a^Z_{l_L}}\in[-0.2,0.1]\, ,\ \  
&\frac{\delta \widehat a^Z_{l_L}}{\widehat a^Z_{l_L}}\in[-8,7]\times 10^{-2}\ ,\ \ 
& b^Z_{l_L}\in[-2,5]\times 10^{-2},\ \ 
& \widehat b^Z_{l_L}\in[-2,5]\ ,\times 10^{-2},\nonumber \\
&\frac{\delta a^Z_{l_R}}{a^Z_{l_R}}\in[-0.2,0.3]\, , \ \ 
&\frac{\delta \widehat a^Z_{l_R}}{\widehat a^Z_{l_R}}\in[-8,7]\times 10^{-2},\ \
&b^Z_{l_R}\in[-3,2]\times 10^{-2},\ \ 
& \widehat b^Z_{l_R}\in[-2,5]\times 10^{-2}\, .\nonumber
\end{align}
Although the allowed range in $a^Z_{l_{L,R}}$ is quite large, we notice
that their  impact on the total amplitude, when summed over lepton chiralities, is much smaller, $2\sum_{l=l_L,l_R}g_Z^{l}a^Z_{l}/\sum_{l=l_L,l_R}(g_Z^{l})^2\in [-6,4]\times 10^{-2}$.

It is interesting to notice that in the limit $g'\to 0$ the result of \eq{asbsobs} is custodial invariant, {\it i.e.}, one finds equal corrections for $Z$ and $W$. This is because, for $g'\to 0$, the only  Wilson coefficients breaking the custodial  symmetry are $c_T$  and $c^f_{L,R}$ \cite{eemp2} that, being     constrained  at the per-mille, have been dropped from \eq{asbsobs}.
We then find that  the test of the custodial symmetry used at LHC \cite{LHCHiggsCrossSectionWorkingGroup:2012nn} 
defined as
\be
\lambda_{WZ}^2\equiv \frac{\Gamma(h\to WW^{(*)})}{\Gamma^{\rm SM}(h\to WW^{(*)})}\frac{\Gamma^{\rm SM}(h\to ZZ^{(*)})}{\Gamma(h\to ZZ^{(*)})}\, ,
\label{custodef}
\ee
is constrained by
\bea
\lambda_{W Z}^2-1
&\simeq& s^2_{\theta_W}\left[0.9c_W-2.6 c_B+3\kappa_{HW}-3.9\kappa_{HB}\right]
\nonumber \\
&\simeq& 0.6 \delta g^Z_1 - 0.5  \delta \kappa_\gamma -1.6 \kappa_{Z\gamma}
 \in [-6,8]\times 10^{-2}\, ,
\label{custo}
\eea
where  the numerical values of the first line have been taken from  \cite{Contino:2013kra}.
We see that    \eq{custo} puts  a bound on $\lambda_{WZ}$  much stronger than
 the present  direct experimental limit given by  \cite{ATLAS:2013sla}~\footnote{The experimental bound on $\lambda_{WZ}$ is extracted 
 not only using \eq{custodef} but also considering  custodial breaking effects in vector-boson fusion. The impact of these latter effects is however negligible.}:
  $(\lambda_{WZ}-1)\in [-0.5,0.1]$.

Along the lines presented here we could 
also study the corrections from Wilson coefficients 
to  the Higgs production modes $\bar ff\to Vh$ and   $VV\to h$ 
that we could similarly show that are constrained by our previous analysis.

%%%%%%%%%%%%%%%%%%%%%%%%%%%%%%%%%%%%%%%%%%%%%%%%%%%%%%%%%%%%%%%%%%%%%%%%%%%%%%%%%%%
\section{CP-odd operators}\label{sec:3}

For completeness, we  show here how 
CP-odd operators enter in TGC and in the process $h\to V\bar f f$, and how they can be  related. 
These operators are \footnote{A  CP-odd operator involving 3 gluon field-strengths and  the operators 
$iy_f|H|^2\bar f_L H f_R+$h.c. complete the list of CP-odd operators; since they do not enter in the observables  discussed here, they have been omitted.}
 \begin{align}
{\cal O}_{B\widetilde B}&={g}^{\prime 2} |H|^2 B_{\mu\nu}\widetilde B^{\mu\nu}\,,
&{\cal O}_{G\widetilde G}&=g_s^2 |H|^2 G_{\mu\nu}^A \widetilde G^{A\mu\nu}\,,\nonumber\\
{\cal O}_{H\widetilde W}&=i g(D^\mu H)^\dagger\sigma^a(D^\nu H)\widetilde W^a_{\mu\nu}\,,
&{\cal O}_{H\widetilde B}&=i g'(D^\mu H)^\dagger(D^\nu H)\widetilde B_{\mu\nu}\,,\\
&&\hspace{-4cm}{\cal O}_{3\widetilde W}=  \frac{1}{3!}g\epsilon_{abc}W^{a\, \nu}_{\mu}W^{b}_{\nu\rho}\widetilde  W^{c\, \rho\mu}\, ,&\nonumber
 \end{align}
where $\widetilde V^{\mu\nu}=\epsilon^{\mu\nu\rho\sigma}V_{\rho\sigma}/2$ for $V=W,B,G$.
These  operators affect TGC as \cite{Hagiwara:1986vm}:
\begin{equation}
\Delta {\cal L}_{3\widetilde V} =
ig \left(  \delta\widetilde \kappa_Z c_{\theta_W} \widetilde Z^{\mu\nu} +\delta\widetilde\kappa_\gamma s_{\theta_W} \widetilde A^{\mu\nu}\right) 
W^-_\mu  W^+_\nu
+
\frac{ig}{m^2_W}\left(  \widetilde\lambda_Z c_{\theta_W} \widetilde Z^{\mu\nu} +\widetilde \lambda_\gamma s_{\theta_W} \widetilde A^{\mu\nu}\right) W^{- \rho}_{\nu}  W^{+}_{\rho \mu}\ ,\nonumber
\label{V3CPO}
\end{equation}
where
\be
  \delta \widetilde \kappa_\gamma = \kappa_{H\widetilde W}+
\kappa_{H\widetilde B}\ ,\qquad
 \delta \widetilde \kappa_Z=- t^2_{\theta_W}  \delta \widetilde \kappa_\gamma\, ,\qquad 
\widetilde  \lambda_Z = \widetilde \lambda_\gamma =  \kappa_{3\widetilde W} \, .
\label{3vertCPO}
\ee
Although the experimental collaborations have not devoted explicit searches for these types of structures, constraints 
on the above coefficients can   be obtained in a near future. For example,
 Ref.~\cite{Dawson:2013owa} finds that with 10 fb$^{-1}$ the 14 TeV  LHC can potentially obtain $|\delta\widetilde\kappa_\gamma|\lesssim 0.06$.

The contribution from CP-odd operators to  Higgs physics can be read from the Lagrangian
\bea
 \Delta{\cal L}_{hV\widetilde V} &= &  - 2\frac{h}{v} \, \Big[c_{W\widetilde W} \, W^{+\mu \nu} \widetilde W^-_{\mu \nu} +
 c_{Z\widetilde Z}\, 
 Z^{\mu \nu}\widetilde Z_{\mu \nu}
 -2t_{\theta_W}\widetilde \kappa_{Z\gamma} \, A^{\mu\nu} \widetilde Z_{\mu \nu}\nonumber \\
 & -&2\kappa_{B\widetilde B}s_{\theta_W}^2A_{\mu\nu}\widetilde A^{\mu\nu}
 -2\kappa_{G\widetilde G}\frac{g_s^2}{g^2}\, G_{\mu\nu}^A\widetilde G^{\mu\nu\, A}
\Big]\ ,
\label{LhVVCPO}
\eea
where, quite analogously to the CP-even case, we have
\be
c_{W\widetilde W}=\, \kappa_{H\widetilde W}\ ,\  \
c_{Z\widetilde Z}=\frac{1}{2} (\kappa_{H\widetilde W} +\kappa_{H\widetilde B} t^2_{\theta_W})-2 \frac{s_{\theta_W}^4}{c_{\theta_W}^2} \kappa_{B\widetilde B}\ ,\ \ 
\widetilde\kappa_{Z\gamma}=\frac{1}{4}(\kappa_{H\widetilde B}-
\kappa_{H\widetilde W})-2s^2_{\theta_W}\kappa_{B\widetilde B}\, .
\label{ctildes}
\ee
Finally, the CP-odd contribution to the $h\to V \bar f f$  amplitude is  given by
\begin{align}
 \mathcal{M}(h \to VJ_{f}) = \,  (\sqrt{2} G_F)^{1/2} \epsilon^{*\mu}(q) \,  J^{V\,\!\nu}_{f}(p) \left[  C^V_f \epsilon_{\mu\nu\alpha\beta}\, p^{\alpha}\, q^{\beta} \right] \, ,
 \end{align}
where we find \cite{MAX}
\be
\ \ \ \ \ \ 
C^V_f =   c_f^V\frac{1}{p^2-m_V^2} + \widehat c_f^{\,V}\frac{1}{p^2}\, \ \ \ \ \ \ \ \ 
 (\widehat c^V_f=0\ {\rm for}\ V=W ) \, ,
\ee
with
\begin{align}
c_f^Z&=  8 g_Z^fc_{Z\widetilde Z}\,,  & c_f^W &=4g_W^f  c_{W\widetilde W} \,, \\
\widehat c_f^{\, Z}&= -8eQ_f t_{\theta_W}   \widetilde\kappa_{Z\gamma}\, . 
 \end{align}
Using  \eq{3vertCPO} and \eq{ctildes}, we  find the relations
\begin{align}
\label{cpoddobs}
c_f^Z&=  \frac{2g_Z^f}{c^2_{\theta_W}} \left(  \delta \widetilde \kappa_\gamma-4c_{2\theta_W}  \widetilde\kappa_{Z\gamma}
-2s^2_{2\theta_W} \kappa_{B\widetilde B}\right)\,,  &
c_f^W &=2g_W^f \left( \delta \widetilde \kappa_\gamma-4  \widetilde\kappa_{Z\gamma}
-8s_{\theta_W}^2 \kappa_{B\widetilde B}\right) \,.
 \end{align}

To our knowledge,  the only CP-odd observables measured at present
are   those in the decay $h\to Z Z^*\to 4l$  \cite{CMS:xwa}, but  only very weak constraints  can  be extracted for the respective Wilson coefficients. 
Nevertheless, the experimental bound on   $BR(h\to Z\gamma)$  and 
  $BR(h\to\gamma\gamma)$ can also be  used to constrain the squared of $\widetilde\kappa_{Z\gamma}$ and  $\kappa_{B\widetilde B}$.   \footnote{The CP-odd and CP-even contributions add  quadratically in the 
  Higgs  branching ratios, giving then  independent bounds on the corresponding  Wilson coefficients.}
  One obtains  similar constraints to their CP-even counterparts,
     since the interference term involves the SM contribution  that is one-loop suppressed. Neglecting 
$\widetilde\kappa_{Z\gamma}$ and  $\kappa_{B\widetilde B}$ in  \eq{cpoddobs}, one obtains a one-to-one relation between the CP-odd contributions to $h\to V \bar f f$ 
and the TGC   parameter $\delta\widetilde \kappa_\gamma$
that can be measured in the near future.

\section{Conclusions}

We have made a first step towards a complete SM fit, 
focusing here  on   EWSB effects in gauge-bosons and Higgs physics.
We have characterized all possible deformations from SM physics  using   the  Wilson coefficients of 
the independent dimension-6 operators, the  relevant ones for our analysis given in Table~\ref{operatorsdim6}.
Assuming flavor-universality (but also including the    top-quark operators separately) 
and  taking as the input data $m_Z$, $G_F$, $\alpha_{\rm em}$ and $m_f$, $m_h$, 
we have grouped   the  operators according to their impact on the   different  experimental data.

\begin{figure}[top]
\begin{center}
\hskip.5cm
\includegraphics[height=6.9cm]{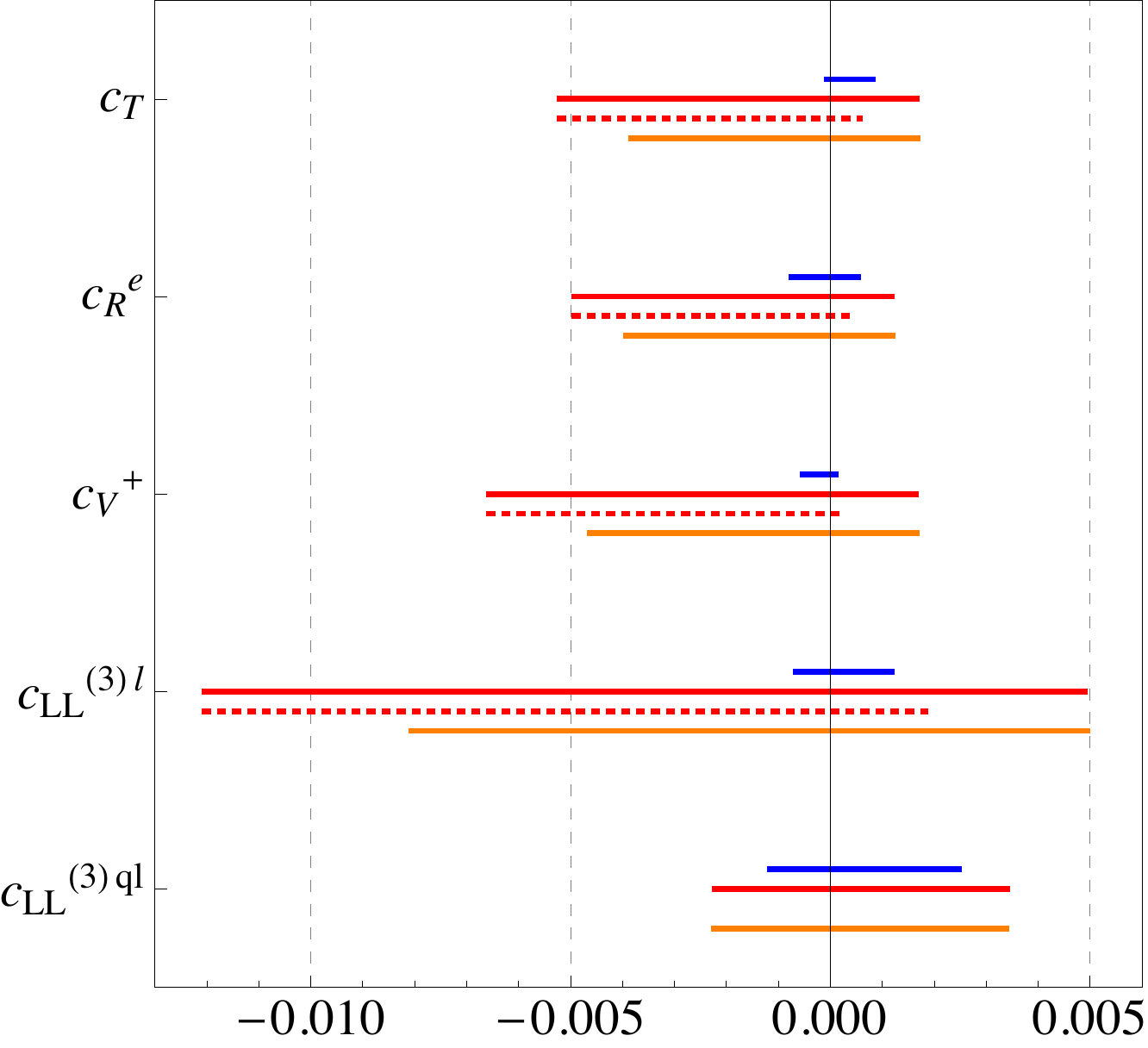}\hspace{5mm}\includegraphics[height=6.9cm]{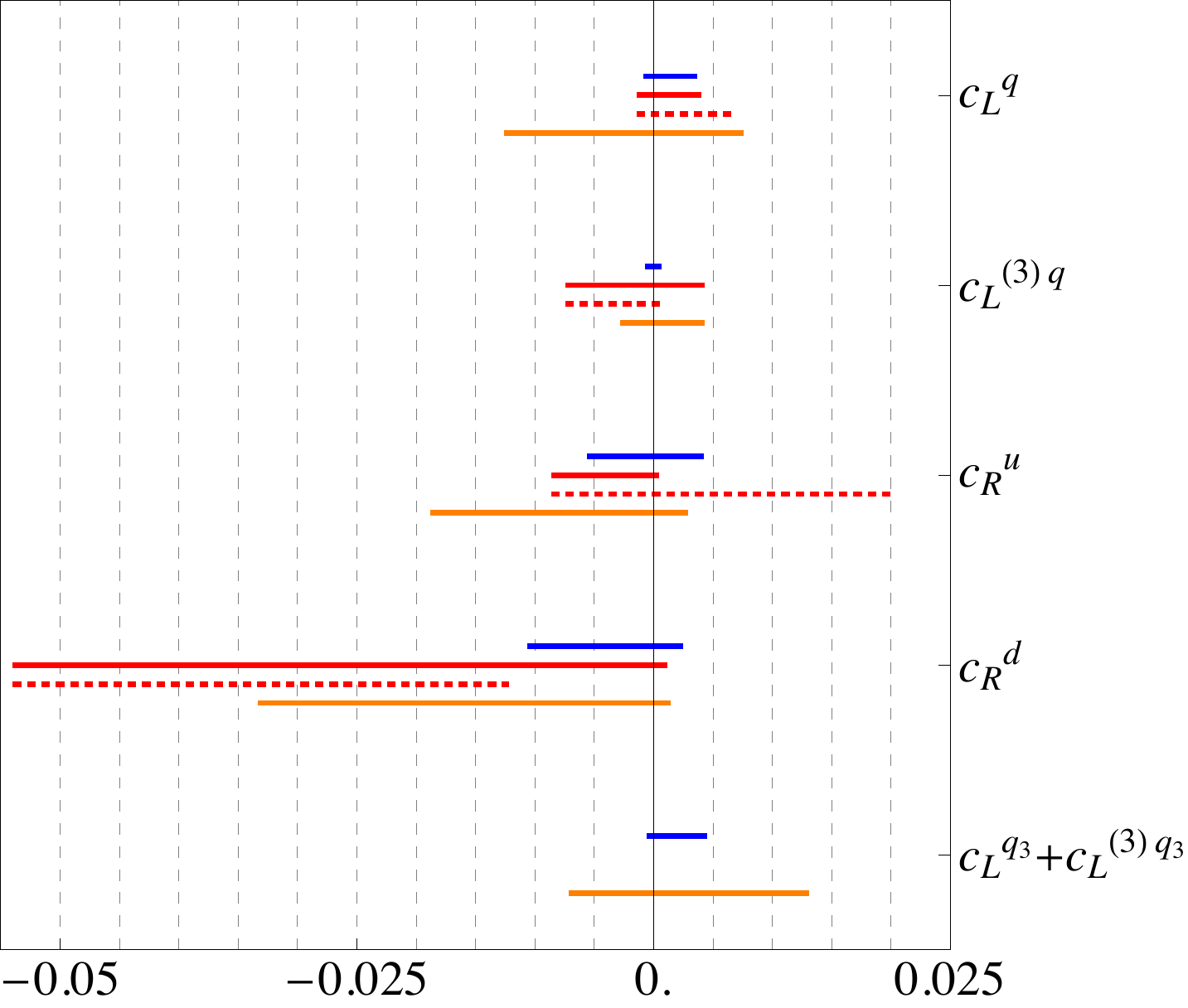}\hspace{5mm}\end{center}
     \caption{\footnotesize \emph{Bounds on the Wilson coefficients of  Eqs.~(\ref{OpsInput}), (\ref{ObsLep1}), (\ref{OpsQuarks1}) and (\ref{opsq3}), from LEP-I, SLC and KLOE data (and the necessary information from $pp\to l\bar\nu$ at the LHC  --see Appendix~\ref{app:cql3}). Solid red lines denote the 95\% C.L. intervals after marginalizing over all other parameters (dashed  lines do not    include KLOE data); blue lines are obtained by setting all other coefficients to zero.
      The global fit including  the operator  of \eq{OpsQuarksMFV}
corresponds to the orange lines.}}\label{fig:P1000}
\end{figure}

In a first group we have the  operators that can affect  the gauge-boson propagators 
and their couplings to fermions.
These  receive strong constraints from well-measured quantities,
mainly  the  $Z$-pole observables at LEP-I and SLC, the $W$ mass at Tevatron, together with the check of the CKM unitarity 
 from low-energy data.
The constraints on the corresponding Wilson coefficients  from a global fit
are summarized  in Fig.~\ref{fig:P1000}.
 To stress the correlation between the various operators, we compare the 95\% C.L. contours obtained by marginalizing all other coefficients (in red) with the contours obtained by setting all other coefficients to zero (in blue). 
More data, such as the low-energy determination of the $\nu$-nucleon and $\nu$-$e$ scattering or atomic parity violation experiments, could  be added to our analysis. Nevertheless, due to their poorer resolution, we do not expect these data to affect substantially our results.

\begin{figure}[top]
\begin{center}
\hskip.4cm
\includegraphics[height=3.9cm]{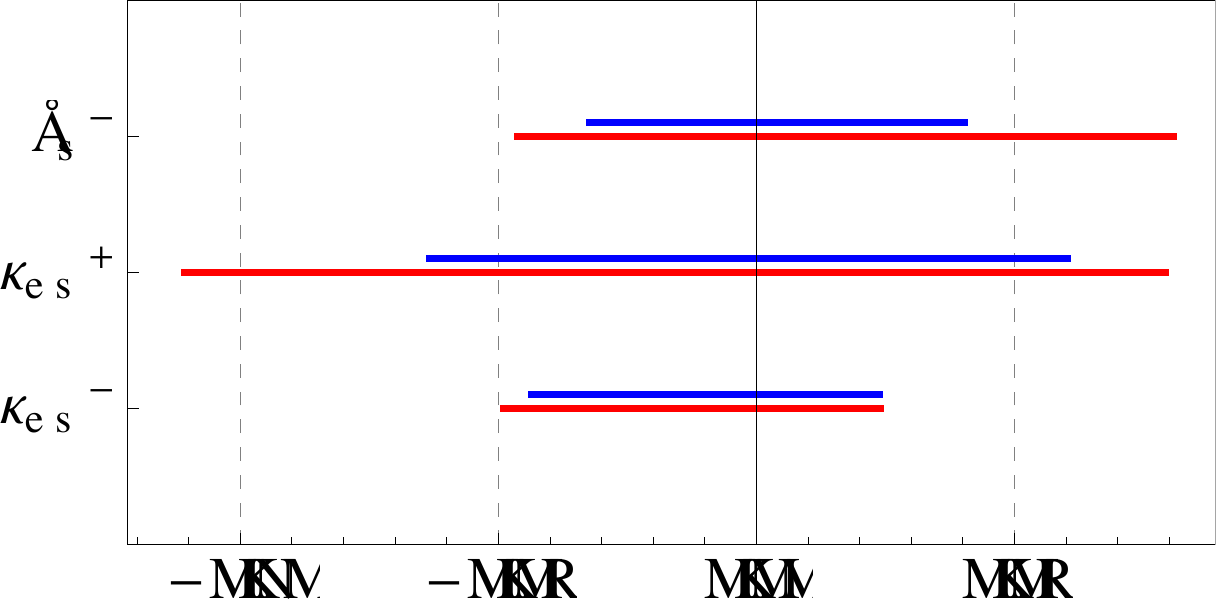}\hspace{4mm}\includegraphics[height=3.9cm]{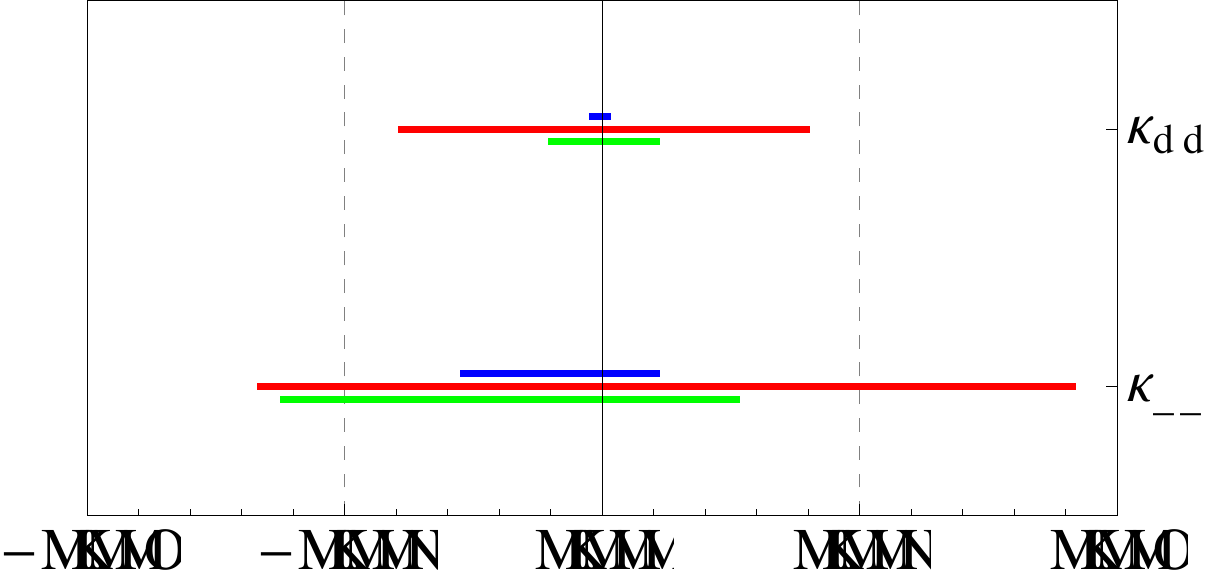}
\end{center}
     \caption{\footnotesize \emph{{Left: Bounds on some of  the Wilson coefficients of \eq{tgc} from LEP-II TGC data and 
     $BR(h\to Z\gamma)$.
Right:    Bounds on some of the Wilson coefficients of \eq{higgsop} from   measurements of the Higgs BR at the LHC. Color code as in 
Fig.~\ref{fig:P1000}, with  green lines  including  the  theoretical prior on the coefficients $c_H$ and $c_{y_{f}}$ that cannot introduce modifications larger than 50\% on  SM predictions.}}}
\label{fig:P100}
\end{figure}

In a second group we have operators affecting triple  gauge-boson vertices, that 
contrary to the previous group, receive milder constraints.
Our basis is suitable for treating  separately these  two groups,  while other bases, such as the one of Ref.~\cite{Grzadkowski:2010es} used in the fit of Ref.~\cite{Grinstein:2013vsa},
makes this separation more difficult,  due to strong correlations between bounds on different Wilson coefficients.  
 We have 3  (combinations of) Wilson coefficients  parametrizing these deviations, given in \eq{tgc}.
Using the 2-parameter fit  of LEP-II  \cite{LEP2wwzOLD} where $\lambda_\gamma$ (and therefore $\kappa_{3W}$)  is neglected, we have presented  bounds on  two of them, \eq{tgcbounds}. Even if we allow for a nonzero  
$\lambda_\gamma$,  LEP-II data is expected to be able to constrain all the 3 coefficients at the per-cent level \cite{Abdallah:2010zj}, but
 a combined three-parameter fit is still not available.
 In  Fig.~\ref{fig:P100} we show our results as bounds on the
 two coefficients
$c^-_{V}$ 
and $ \kappa^+_{HV}$ (once the data on $BR(h\to Z\gamma)$ is also used to constrain the coefficient  $\kappa^-_{HV}$).

Finally, 
there are the  operators that only affect Higgs physics.
These are  8 CP-even operators (for a single family)  that
can  affect, independently, the Higgs $BR$ to fermions, $\gamma\gamma$ and $Z\gamma$,
as well as  production cross-sections both in the gluon-gluon  and vector-boson fusion modes
 (and also the triple-Higgs coupling).
In Fig.~\ref{fig:P100} 
we  present the  model-independent constraints on  the Wilson coefficients entering in the  $h\gamma\gamma$ and  $hGG$ effective-couplings. These  are very severe bounds   as the BSM contributions enter at the tree-level and can then compete with the  SM contributions arising at the one-loop level.
Constraints on other Wilson coefficients, $c_H$ and $c_{y_f}$, are at present very mild and therefore not shown.
The Wilson coefficient $\kappa^-_{HV}$  enters in $BR(h\to Z\gamma)$
and is constrained, as we said, only form
the present experimental bound  $BR(h\to Z\gamma)\lesssim 10\times BR(h\to Z\gamma)_{\rm SM}$ \cite{Chatrchyan:2013vaa,ATLAS:2013rma}, leaving still room for large deviations
 from the SM.

Having determined all Wilson coefficients, any other (Higgs)  process 
will depend on  the above physics and therefore their  BSM effects can be indirectly constrained.
In particular,
we have studied  $h\to V\bar f f$.
Neglecting corrections tightly bounded from the first group of observables, 
we have shown that  the different form-factors of the $h\to V\bar f f$ amplitude  
are related with  TGC and  $BR(h\to Z\gamma)$.
These latter are  at present  already constrained, and 
in the future LHC can considerably improve these bounds,
being   more competitive than looking for deviations in the $h\to V\bar f f$   decays.
 Similar arguments apply to CP-odd operators. 
 
  Possible deviations in the approximate relations derived here, \eq{asbsobs}, \eq{custo} and \eq{cpoddobs},
  would imply the breakdown of our assumptions,  hinting possibly  towards a non-linear realization of EWSB, in which $h$ is not part of a Higgs doublet and no connection between Higgs and gauge-boson physics can be made.

%%%%%%%%%%%%%%%
\section*{Acknowledgements}
%%%%%%%%%%%%%%
We are particularly grateful to M.~Montull for collaboration in the early phase of this project. We also thank M.~Carrer, J.~Elias-Miro, J.R.~Espinosa, E.~Masso for discussions. FR thanks A.~Avetisyan, O.~Matsedonskii, A.~Thamm and in particular K.~Kotov for discussions about the analysis of Appendix~\ref{app:cql3}. The work of AP was partly supported by the projects FPA2011-25948, 2009SGR894 and ICREA Academia Program, while FR  acknowledges support from the Swiss National Science Foundation, under the Ambizione grant PZ00P2 136932 and thanks IFAE, Barcelona, for hospitality. We are grateful to the Galileo Galilei Institute for Theoretical Physics for the hospitality and the INFN for partial support during the completion of this work.

%%%%%%%%%%%%%%%%%%%%%%%%%%%%%%%%%%%%%%%%%%%%%%%%%%%%%%%%%%%%%%%%%%%%%%%%%%%%%%%%%%%%%%%%%%%%%%%%%%%%%%%%%%%%%%%%%%%%%%%%%%%%%%%%%%%%%%%%%%%%%%%%%%%%%%
\appendix
%%%%%%%%%%%%%%%%%%%%%%%%%%%%%%
\section{Corrections to $V\bar f f$ couplings and $Z$-pole observables}
\label{app:LEP}
The couplings between the $Z$-boson and  fermions are altered as
\begin{equation}
g_Z^f=g_Z^{f,\rm SM}+\delta g_Z^f+\delta g_Z^{f,{\rm uni}}\, ,
\end{equation}
where 
\begin{equation}
g_Z^{f, \rm SM}=m_Z\left(\sqrt{2}G_F\right)^{1/2}\left(T^3_{L\, f}-Q_fs^2_{\theta_W}\right)\, ,
\end{equation}
is the SM value, $T^3_{L\, f}$ and $Q_f$ are respectively  the weak-isospin and charge  of the fermion $f=\{f_L,f_R\}$, and $s^2_{\theta_W}\equiv(1-\sqrt{1-2^{3/2}\pi\alpha_{\rm em} /m_Z^2G_F})/2$. We have divided the corrections into a fermion-specific part $\delta g_Z^{f}$,
\begin{equation}
\delta g_Z^{f_L}=m_Z\left(\sqrt{2}G_F\right)^{1/2} \left(T^3_{L,f}c_L^{(3)\,f}-\frac{c_L^f}{2}\right)\, ,\quad\quad
\delta g_Z^{f_R}=-m_Z\left(\sqrt{2}G_F\right)^{1/2}\frac{c_R^f}{2}\, ,
\label{shifts}
\end{equation}
(notice that for leptons $\delta g_Z^{l_L}=0$ in our basis) and a universal part,
 \begin{equation}\label{shiftUNI}
\delta g_Z^{f,{\rm uni}}= -m_Z\left(\sqrt{2}G_F\right)^{1/2}Q_f\, \delta s^2_{\theta_W}+\frac{g_Z^{f}}{2}\left(\hat T-\frac{\delta G_F}{G_F} \right),
\end{equation}
which includes modifications to the wave-function of the $Z$-boson, $\delta\Pi_Z^\prime(m_Z^2)=2\hat S s^2_{\theta_W}$, corrections to the input parameters, and contributions to $\Pi_{\gamma Z}(m_Z^2)$ through
\begin{equation}
\delta s^2_{\theta_W}= \frac{1}{c_{2\theta_W}}\left[s^2_{\theta_W}\hat S -\frac{s^2_{2\theta_W}}{4}\left(\hat T-\frac{\delta G_F}{G_F}\right)\right]\, .
\label{deltas2W}
\end{equation}
This is enough to compute  the corrections to   the observables that we use in the fit (the uncorrelated subsets $\{A_l,R_l, \sigma_{\rm had} ^0,\Gamma_Z\}$ and $\{R_b, R_c, A^{0,b}_{FB},A^{0,c}_{FB}, A_b, A_c\}$ from Ref.~\cite{ALEPH:2005ab}):
\begin{equation}\label{LEPobservables}
{ A_{l}} \equiv {(g_Z^{l_L})^2 - (g_Z^{l_R})^2 \over (g_Z^{l_L})^2 + (g_Z^{l_R})^2 } \ ,\quad { R_l} \equiv {\Gamma_Z^{\rm had}  \over \Gamma_Z^{l} }\ ,\quad { \sigma_{\rm had} ^0}\equiv {12 \pi \over  m_Z^2} {\Gamma^{l}_Z \Gamma^{\rm had}_Z \over \Gamma_Z^2} \, , 
\end{equation}
and 
\begin{gather}
{R_q}=\frac{(g_Z^{q_L})^2+(g_Z^{q_R})^2}{3 [(g_Z^{d_L})^2 +(g_Z^{d_R})^2] + 2 [(g_Z^{u_L})^2 + (g_Z^{u_R})^2]}
,\quad { A_q}=\frac{(g_Z^{q_L})^2-(g_Z^{q_R})^2}{(g_Z^{q_L})^2+(g_Z^{q_R})^2},\quad { A_{FB}^{0,q}}=\frac{3}{4}A_qA_l\,,
\end{gather}
where $q=b,c$ at LEP. The partial widths are defined as,
\begin{gather}
\Gamma_Z^{l}={  m_Z  \over 24 \pi}[(g_Z^{l_L})^2 + (g_Z^{l_R})^2] ,\quad\Gamma_Z^{\rm had}= {  m_Z  \over 24 \pi}\left(3N_c [(g_Z^{d_L})^2 +(g_Z^{d_R})^2] + 2 N_c[(g_Z^{u_L})^2 + (g_Z^{u_R})^2]  \right)\,,\nonumber\\
 \Gamma_Z^{\nu}={  m_Z  \over 24 \pi}  (g_Z^{\nu})^2,\quad 
\Gamma_Z=\Gamma_Z^{\rm had}+3\Gamma_Z^{l}+3\Gamma_Z^{\nu}\, .
\end{gather}
Notice that the universal corrections due to $\delta\Pi_Z^\prime(m_Z^2)$ cancel out from the observables \eq{LEPobservables}, but enter in the total width (and similarly for the partial widths) which is corrected as
\begin{equation}\label{shiftGammaZtot}
{\frac{\delta \Gamma_Z}{\Gamma_Z}}=
2\frac{\sum_f  g_Z^{f}(\delta g_Z^{f,\textrm{uni}}+\delta g_Z^{f})}{\sum_f  (g_Z^{f})^2}\, .
\end{equation}
Finally, the couplings of $W$-bosons to fermions are modified as
\begin{equation}
g_W^f=g_W^{f,\rm SM}+\delta g_W^f+\delta g_W^{f,\rm uni}\, ,
\end{equation}
where 
\begin{equation}\label{deltashiftW}
g_W^{f,\rm SM}=\frac{m_W G_F^{1/2}}{2^{1/4}}\, , \quad\quad \frac{\delta g_W^{f}}{g_W^{f}}=c_L^{(3)\,f}\,,\quad\quad
  \frac{\delta g_W^{f,\rm uni}}{g_W^{f}}=\frac{\delta m_W}{m_W}-\frac{\delta G_F}{2G_F}\, .
\end{equation}

%%%%%%%%%%%%%%%%%%%%%%%%%%%%%%%%%%%%%%%%%%%%%%%%%%%%%%%%%%
\section{$q\bar q'  l\bar\nu$ contact-interactions at the LHC}
\label{app:cql3}

Since the fermion-fermion scattering amplitude mediated by  contact-interactions grows with energy, these can be tested with accuracy at the LHC.  We are interested here to put a bound on  the 4-fermion operator  $\mathcal{O}_{LL}^{(3)\, ql}$.
This  operator  affects in particular   the cross-section of  $pp\to l\bar  \nu$ that has been measured
at the LHC.
Nevertheless, the only LHC analysis  \cite{CMSHNC} has been on  the 
 helicity non-conserving (HNC) operator $(Q_L^r\bar u_R)\epsilon_{rs}(L_L^s \bar e_R)$,   which also  modifies
  $pp\to l\bar \nu$.
Contrary to  $\mathcal{O}_{LL}^{(3)\, ql}$,  however, the HNC operator does not interfere with the SM contribution,
so that the results of Ref.~\cite{CMSHNC} cannot be used for 
$\mathcal{O}_{LL}^{(3)\, ql}$ and a dedicated analysis is necessary.

\begin{figure}[top]
\begin{center}
\includegraphics[height=3.9cm]{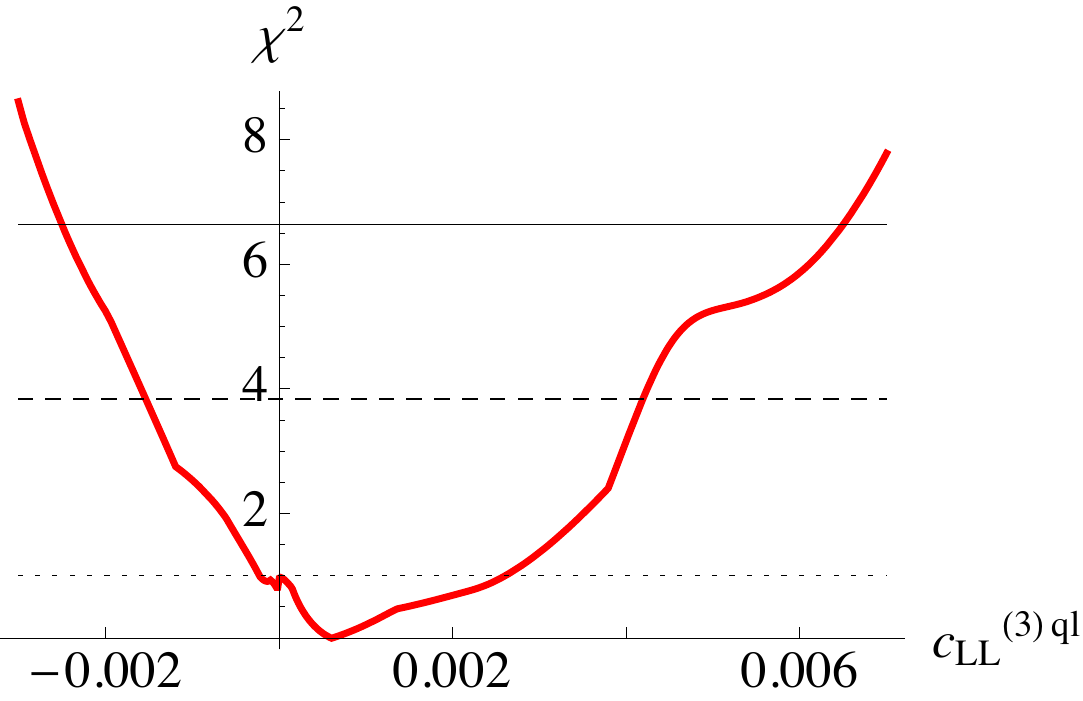}
\end{center}
     \caption{\footnotesize \emph{The $\chi^2$-distribution for the Wilson coefficient of the operator $\mathcal{O}_{LL}^{(3)\, ql}$. }}\label{fig:chi2}
\end{figure}

A signal sample needs to be considered for every value of $c_{LL}^{(3)\, ql}$ and then compared with the data.
 Other operators entering in   $pp\to l\bar \nu$, such as the HNC operator,
 are suppressed by small Yukawa couplings under the MFV assumption, that we consider here, and can be neglected.
 We simulate the effects of $\mathcal{O}_{LL}^{(3)\, ql}$ by integrating out a heavy $W'$-boson, implemented with FeynRules\cite{Christensen:2008py} and generate parton-level events with MadGraph~\cite{Alwall:2011uj}. 
We divide the signal and the data in 3 regions, according to the transverse mass of electron and neutrino:  $M_T\in[1,1.5]$ TeV, $M_T\in[1.5,2]$ TeV  and $M_T>2$ TeV,  which we treat as statistically independent. Ref.~\cite{CMSHNC} reports, respectively, the observation of 22, 0 and 1 events in these regions. 
We compute C.L. contours, assuming a Poissonian distribution around the signal+background and treating errors as nuisances (an estimated 5\% systematic error in the signal plus a 4.4\% in the luminosity are summed in quadrature). We checked that such estimates reproduced satisfactorily the results in the case of the HNC model. Our results, in form of the $\chi^2$-distribution, are summarized in Fig.~\ref{fig:chi2} and imply
\begin{equation}\label{boundclq3}
-0.001 \lesssim c_{LL}^{(3)\, ql}\lesssim 0.004 \quad \textrm{at} \,\,95\%\,\textrm{C.L.}\,\,.
\end{equation}
Combining it with the muon channel can give   better limits in flavour-universal  BSM models,  as the ones considered here. Notice that bounds on $c_{LL}^{(3)\, ql}$  from the differential distribution of  $pp\to l^+l^-$ at the LHC \cite{deBlas:2013qqa} are weakened once the contributions from other operators are taken into account. Therefore, \eq{boundclq3} provides at present the best model-independent bound on this Wilson coefficient.

\end{document}